\documentclass[%
 aip,
rsi,%
 amsmath,amssymb,amsfonts,
 reprint,%
]{revtex4-1}

\usepackage{graphicx}
\usepackage{dcolumn}
\usepackage{bm}

\usepackage{color}
\definecolor{red}{rgb}{1,0,0}

\definecolor{blue}{rgb}{0,0,1}

\begin{document}

\title{Space-time nature of causality}
%



\author{Ezequiel Bianco$-$Martinez}
\affiliation{Institute of Complex Sciences and Mathematical Biology, University of Aberdeen, SUPA, AB24 3UE, Aberdeen, UK}
\affiliation{ZorgDomein, Straatweg 68, 3621BR, Breukelen, Netherlands}

\author{Murilo S. Baptista}
\email[]{murilo.baptista@abdn.ac.uk}
\affiliation{Institute of Complex Sciences and Mathematical Biology, University of Aberdeen, SUPA, AB24 3UE, Aberdeen, UK}

\date{\today}

\begin{abstract}
In a causal world the direction of the time arrow dictates how past causal events in a variable $X$ produce future effects in $Y$. $X$ is said to cause an effect in $Y$, if 
the predictability (uncertainty) about the future states of $Y$ increases (decreases) as its own past and the past of $X$ are taken into consideration. Causality is thus intrinsic dependent on the observation of the past events of both variables involved, to the prediction (or uncertainty reduction) of future event of the other variable.     
We will show that this {{temporal}} notion of causality leads to another natural {{spatio-temporal}} definition for it, 
{{and that}}  can be exploited to detect the arrow of influence from $X$ to $Y$, either by considering shorter time-series of $X$ and longer time-series of $Y$ (an approach that explores the time nature of causality) or  lower precision measured time-series in $X$ and higher precision measured time-series in $Y$ (an approach that explores the spatial nature of causality).  Causality has thus space and time signatures, causing a break of symmetry in the topology of the probabilistic space, or causing a  break of symmetry in the length of the measured time-series, a consequence of the fact that information flows from $X$ to $Y$.
\end{abstract}
\maketitle

\section*{Introduction}

{\bf In a causal world the direction of the time arrow dictates how past causal events produce future effects. The determination of the direction and the intensity of the arrow of influence, causality, is one of the first questions one tries to answer in order to model a system. In ecology, it is fundamental to understand  whether zooplankton concentration drives fish population. In meteorology, one wishes to determine whether and how surface sea temperature affects atmospheric temperature in different parts of the globe, or how green house gases drive global temperature. In finance, tax  and expenditure correlates with saving and growth. In geology, one wants to access the direction of the flows of underground water from some measurements of water reservoir levels. In urbanism, one wants to understand how electricity consumption drives (or is driven by) urbanism or how building environment leads to obesity. Given the relevance of the topic, several methods have been developed in the last decades to study causality. Among then, there are the approaches that access causality based on informational quantities. They are sustained by the fundamental idea that if $X$ causes an effect in $Y$, then uncertainty about future states of $Y$ is reduced by considering the past of $Y$ and the past of $X$, a hypothesis that implicitly adopts the Granger causal idea that observations in the past of both $X$ (causing system) and $Y$ (where the effect is produced) can be used to predict the future state of $Y$. This work aims at unifying the Granger definition of causality defined in terms of predictability with those based on information quantities by studying the spatio-temporal dynamics of causality. We will show that {{if}} a system $X$ causes an effect in a system $Y$, {{then not only causal information from $X$ to $Y$ is positive, but also longer-time or higher-resolution}} observations in $Y$ can be used to predict the past states of the system $X$, an observation that will lead us to propose a new informational theoretic quantity that we name Causal Mutual Information (CaMI), {{and that can assist us in easily quantifying the direction of the flow of information. This work will}} show that causality has space and time signatures, and each signature can be advantageously exploited to study the direction of influence in different systems. Moreover, we will show that our quantity allows for a simple, experimental appealing and less computational demanding approach, but rigorous, quantification of causality.}  

The determination that a past event in a system has caused a present effect on another system provides a straight measurement of the direction of influence in these systems. Causal relationships between two events happening in two different systems $X$ and $Y$ can be established by verifying whether past events in $X$ and $Y$ influence future events in $Y$. Such understanding is fundamental to characterise, model, and predict behaviour in natural, social, and technological systems. The study of the cause-effect relationships is defined as causality. Causality is a concept that envolves the temporal relationship among past, present and future events of variables. Our studies show however that the temporal nature of causality from $X$ to $Y$ can not only be redefined in terms of the reduction of uncertainty from the variable $X$ solely based on observations in the past, present and future, of $Y$ {{(a property that can be understood from the way Transfer Entropy (TE) \cite{schreiber} is defined, and also intuitively derived from Granger causality defined by the way past dynamical states directly and linearly influences dynamical future states)}}, but also that causality can be defined in terms of the topological feature of the probabilistic space. In a deterministic system, two temporally related events defined by two particular states of the system are also intrinsically related in space. This space-time ergodic duality in a deterministic system indicates that time-causality should lead to space-causality, a property that we will explore in this work to create novel ways to quantify causality. {{If there is a direct flow of information from $X$ to $Y$, this is physically interpreted as to that the uncertainty about past of $X$ is reduced by observations of future states of $Y$. 
This physical interpretation of causal information, and that is the core behind the definition of Granger causality, can be demonstrated by analysing the topology of the probabilistic space - exploring the spatio character of causality, without the need to construct a model of the observed data, as it would be the usual procedure from the Granger approach.}}    

When a perturbation affects a system, its influence is transmitted from the perturbation's  source to the other variables of the system. The path the perturbation takes to propagate within the system can be predicted analysing the causality of events in the system. A smart way to study causality is through a controlled experiment where perturbations can be designed to extract the causal structure of the system. However, the desired experiment could be too expensive, technically impossible to perform, or too invasive. Therefore, it is important to develop methods to identify the causal structure of a system only from observational data, without employing any perturbative technique. The identification of the causing and affected systems from observational data has been of great interest for many scientists. Consequently, and also because the identification of causality is 
fundamental to the effective observation, modelling, and controlling of any complex system,  several techniques to infer and quantify 
causality have been recently proposed \cite{mooij,guyon,kano,granger,schreiber,palus,hirata,haufe,chen,ancona,bollt}.

Granger \cite{granger,granger1980testing,granger1988some} considered that if a  variable $x(t)$ causes an effect in $y(t)$, then 
predictions of $y(t)$ are improved  by considering its own past complemented by the past of $x(t)$.
Based on this assumption, he constructed statistical tests to validate this hypothesis. To adopt this hypothesis to study causality from 
data by constructing linear models, Granger causality introduce measures
mostly used in correlation based approaches (directed partial coherence) \cite{amblard2011directed}, 
Directed Coherence \cite{takigawa}, Partial Direct Coherence \cite{baccala},  direct Directed Transfer Function \cite{kaminski}, which are capable of identifying interactions in linear systems, but are not suitable to detect causality among subsystems composing a non-linear system \cite{haufe}. Due to this, methods based on Granger causality appropriated to detect causality in nonlinear system were developed \cite{chen,ancona}.  Granger causality can also be adopted by informational theoretical quantities, of special interest to us, such as the TE, Directed Information Theory \cite{marko1973bidirectional,amblard2011directed}, Conditional Mutual Information \cite{palus}, Partial Transfer Entropy \cite{vakorin}, and Mutual Information from Mixed Embeddings (MIME) \cite{vlachos}, which explore the intuitive notion that if system $X$ causes an effect in system $Y$ then, as specifically defined by transfer entropy, 
the amount of uncertainty in future values of Y is reduced by knowing the past values of X given past values of Y. 

In this work, along the lines of the work in Ref. \cite{amblard2011directed}, we intend to unify the concept of causality based on the predictability {{of dynamical states}} introduced by Granger with the concept of causality defined in terms of transfer entropy, {{ by considering the spatio-temporal character of causality. We will also introduce an}} unnoticed informational quantity that fully explores the space-time properties of causality, and that we call Causal Mutual Information (CaMI).  CaMI$_{X \to Y}$ measures the total amount of information being transmitted from $X$ to $Y$, including both the information shared between both variables and that can be used to predict the present state of $X$ by observations in $Y$ (i.e., Mutual Information), and the causal information transmitted from $X$ to $Y$ and that can be used to predict the past states of $X$ by observations of the past and future states of $Y$ (i.e., Transfer Entropy).   

Our proposed definition of causality, quantified by the quantity CaMI, is based on the physical notion that if $X$ causes effect in $Y$, then longer observations of the variable $Y$ (or alternatively  higher resolution observations) than the one considered for $X$ can be used to predict the past of $X$. CaMI is calculated only by the probabilities of joint events, without the need of conditional events. This allows one to do a reduction in the dimensionality of the probability space used to quantify causality, resulting in a method that demands low computational power, and therefore allows for a quick assessment of causality. {{To illustrate why this is of any interest, notice that when calculating TE one needs to calculate conditional probabilities. However, conditional probabilities require the calculation of joint probabilities. CaMI only considers joint probabilities, sparing one from the need to further calculate conditional probabilities, but nevertheless a quantity that trivially provides the directionality index, the net transfer entropy between two systems. Thus, permitting the study of causality with less computational resources.}} Then, we show that the topology of the probabilistic space {{of joint probabilities}} determined by the shapes and forms of the partitions being generated by a dynamical process can also be used to {{not only visualize the spatio-temporal character of causality but also quantify causality, through the here defined CaMI. The emergence of the spatio-temporal nature of causality can be cumbersome or impossible to obtain with the usual higher-dimensional space of conditional probabilities, from which TE is calculated. A direct application of the topological properties of the joint probabilistic space is that it is not only possible to state about the direction of information between the measured variables but also to determine regions, the here  called causal bubbles, that define ranges for the variables  that are responsible for most of the information transmitted  between two systems. The topology of this space can also be used to demonstrate that if there is a flow of information  from the variable $X$ to $Y$, then it is also true that observations in Y allows for an accurate prediction of location of the  past dynamical states of variable $X$. As we shall see, there are preferential places and preferential times to measure the information being transmitted.  Another advantage of our approach is that our detection of causality is oriented to treat experimental systems, since our probabilistic space is based on partitions, which can be constructed {{based on the available}} experimental resolution of the data, {{or on the sampling rate of the measurements, allowing one to work with longer or short time-series. The usual causal analysis based on the value of TE calculated over equal-sized cells or from probabilities estimated by kernels would only provide a scalar number, with no information about the dynamics behind the process or the topology of the probabilistic space.}} 

\section{Coupled maps}\label{sec:causal_maps}
	 
For this study of causality, we consider discrete coupled maps, whose connected nodes are described by:
\begin{equation}\label{coupled_maps_network}
x^{i}_{n+1}=f(x^{i}_{n},r)(1-\alpha)+\frac{\alpha}{k_{i}}\sum_{j=1}^M A_{ij}f(x^{j}_{n},r),
\end{equation}
where $x_n^{i}$ is the trajectory of map $i$, $n$ is the time index of the variable of the dynamics, $\alpha\in[0,1]$ is the coupling parameter, $A_{ij}$ is the adjacency matrix (with entries of 1 or 0 depending on the existence of a connection between two nodes or not, respectively), $r$ is the fixed parameter of each map, $k_{i}$ is the degree of node $i$ ($k_i=\sum(A_{ij})$ and $f(x_{n},r)$ is the map governing the dynamics that can be described by the Logistic map  
$f_1(x_{n},r)=rx_{n}(1-x_{n})$ {\color{black}\cite{myrberg1962}}. A disconnected node ($k_i=0$) is described by 
$x^{i}_{n+1}=f(x^{i}_{n},r)$.  
We assume there are $N_d$ maps forming the network. 

Giving two variables $X$ and $Y,$ we are interested in determining the direction of influence that one variable imposes over the other one. If $X$ influences $Y,$ we represent this interaction by $X \to Y$.

\section{Partitions, state and probabilistic space, and symbolic trajectories}\label{sec:causal2}

In what follows, we consider that marginal observations are being made in the relevant variables, defining events in one variable by the falling of a trajectory point within an interval. This interval represents the resolution of the observer.  These marginal observations  and their probabilities, will be used to calculate the probabilities to quantify causality. For simplicity in our analysis, we encode these partitions into symbols, and treat the trajectories as symbolic trajectories.

\subsection{Order-$m$ Partitions, symbolic representation, and dynamics on it}

Consider two discrete scalar time-series $\mathbf{X}=\{x_0,x_1,x_2,\cdots,x_{n-1}\}$ and $\mathbf{Y}=\{y_0,y_1,y_2,\cdots,y_{n-1}\}$ with $n$ elements, and $\bf{\Phi}=\{\bf{X}, \bf{Y}\}$ defines a pair of variables taken from two subsystem in a complex network or coupled system.
Therefore, a point in a $2$-dimensional state space $\Omega^{XY}$ with coordinates $[\bf{X} \times \bf{Y}]$ representing the states of the subsystems $\bf{\Phi}$ at time $t$ has coordinates $\{x_{t},y_{t}\}$. 

We define a marginal partition of order-$m$ of the coordinate $X$ as $\mathcal{C}^X(m)$, defined by the boundary curves $\mathcal{L}_X(m)=\{l^X_1(m),\cdots,\l^X_r(m)\}$, which in this work are assumed to be straight lines, orthogonal to the direction of $X$. Then, this partition is composed by columns $\mathfrak{c}^X_i(m)$ where each one is separated from any other by one and only one curve $l^X_i(m)\in \mathcal{L}_X(m)$.
Similarly, for the coordinate $\bf{Y}$ we can define a marginal partition $\mathcal{C}^Y(m)$, formed by rows $\mathfrak{c}_i^Y(m)$, enclosed by the set of boundary curves $\mathcal{L}_Y(m)=\{l^Y_1(m),\cdots,\l^Y_r(m)\}$, which in this work are assumed to be straight lines, orthogonal to the direction of $Y$.
Since we have a 2D time-series, we can construct a space partition $\mathcal{C}^{XY}(m)$ as a splitting of the space $\Omega^{XY}$ formed by the union of the lines in $\mathcal{L}_X(m)$ and $\mathcal{L}_Y(m)$, so 

\begin{equation}
\mathcal{C}^{XY}(m)=\mathcal{C}(m)^X \cup  \mathcal{C}(m)^Y.
\end{equation}

Areas enclosed by the straight lines of $\mathcal{C}^{XY}(m)$ form the cells $\mathfrak{c}^{XY}_i(m)$ of the partition $\mathcal{C}^{XY}(m)$ that are encoded by the symbols $\mathfrak{s}^{XY}_i(m)$. $S^{XY}(m)$ represents all the possible symbols encoding cells in the partition of order-$m$.

The dynamics of points in this partition are represented by the transformation $\mathcal{U}_t$: $(x_{i+1},y_{i+1}) =\mathcal{U}_t(x_i,y_j)$, and $\mathcal{U}_t^p(x_t,y_t)=(x_{t+p},y_{t+p})$.  The symbolic dynamics of points in this partition are regulated by the transformation $\mathcal{T}$, a surjective mapping of the states of variables in $\Omega^{XY}$ to a specific symbol in $S^{XY}(m)$. $\mathcal{T}$ provides a symbolic sequence $\bf{\Phi}$ in the partition $\mathcal{C}^{XY}(m)$. From Eq. (\ref{coupled_maps_network}), $\mathcal{T}$ is the transformation that maps points from  $\Omega^{XY}$ into itself, a 2D projection of the whole $N_d$-dimensional network. 

Given the partition $\mathcal{C}^{XY}(m)$, we define a transition matrix $\Pi(m)$ where the element $\Pi(m)_{ij}=1$ if the cell $\mathfrak{c}^{XY}_i(m)$ is the pre-image of the cell $\mathfrak{c}^{XY}_j(m)$ (i.e., there is a dynamical evolution from cell $\mathfrak{c}^{XY}_{i}$ to cell $\mathfrak{c}^{XY}_{j}$).

We define a \textit{transition matrix of order-$m$ ($\Pi\wp(m)$)} as:
\begin{equation}
\Pi\wp_{ij}(m) = \left\{ 
  \begin{array}{ll}
    1 			& 			\quad \text{if $F^m(\mathfrak{c}_{i})\cap \mathfrak{c}_{j} \neq \emptyset$}\\
    0           & 			\quad \text{Otherwise}
  \end{array} \right.
\end{equation}

A partition is defined as an order-$m$ if it generates a transition matrix of order $m$. 

We adopt a partition defined by marginal probabilities because we want to define informational measures that quantify the predictability one has to predict the state of one variable by measuring only the state of the other variable, assuming that variables are being measured by a physical process, i.e., there is a measurement resolution.   

\subsection{Probabilistic space and symbolic trajectory}

Now, let us define a $L$ time-delay and time-forward coordinate system from which probabilities are calculated. The time-delay trajectory $\mathbf{\Phi}_{-L}(t)=\{X_{-L}(t), Y_{-L}(t)\} = \{{x_{t-L}, \cdots,x_{t-1}},{y_{t-L},\cdots,y_{t-1}}\}$ represents a short segment with length $L$ (e.g. $L$ points) of the time-series $\bf{\Phi}(t)$ taken for a time spam between the integer time $t-L$ until the time $t-1$, the time $t$ representing the time moment from where past and future are defined. By applying the transformation $\mathcal{T}$ to a segment of length $L$ of the time-series $\mathbf{\Phi}_{-L}(t)$, we generate a sequence of symbols that represent the itinerary followed by the past length-$L$ trajectory. The trajectory 
points $\mathbf{\Phi}_{-L}^{\sigma}(t)$ follow an itinerary along the partitions $\mathcal{C}^{X}(m)$ ($\sigma \equiv X$), $\mathcal{C}^{Y}(m)$ ($\sigma \equiv Y$), or $\mathcal{C}^{XY}(m)$($\sigma \equiv XY$), which are given by $\{ \mathfrak{c}^\sigma_{i_{t-L}}(m),\mathfrak{c}^\sigma_{i_{t-L+1}}(m),\cdots,\mathfrak{c}^\sigma_{i_{t-1}}(m)\}$. If $\sigma=XY$, then $(x_{t-L},y_{t-L}) \in \mathfrak{c}^{XY}_{i_{t-L}}(m)$. If $\sigma=X$, then $x_{t-L} \in \mathfrak{c}^X_{i_{t-L}}(m)$. 
The itinerary $\mathcal{C}_{-L}^\sigma(t,m)=\{\mathfrak{c}^\sigma_{i_{t-L}}(m),\mathfrak{c}^\sigma_{i_{t-L+1}}(m),\cdots,\mathfrak{c}^\sigma_{i_{t-1}}(m)\}$ is encoded by a symbolic sequence $S_{-L}^\sigma(t,m)=\{ \mathfrak{s}^\sigma_{i_{t-L}}(m),\mathfrak{s}^\sigma_{i_{t-L+1}}(m),\cdots,\mathfrak{s}^\sigma_{i_{t-1}}(m)\}$, from which probabilities can be calculated. Similarly, the forward-time trajectory $\mathbf{\Phi}_{L}^\sigma(t)=\{X_L(t),Y_L(t)\}$ follows an itinerary (or visits the sequence of cells) $\mathcal{C}_{L}^\sigma(t,m)=\{\mathfrak{c}^\sigma_{i_{t+1}}(m),\mathfrak{c}^\sigma_{i_{t+2}}(m),\cdots,\mathfrak{c}^\sigma_{i_{t+L-1}}(m)\}$ that is encoded by the symbolic sequence $S_{L}^\sigma(t,m)=\{\mathfrak{s}^\sigma_{i_{t+1}}(m),\mathfrak{s}^\sigma_{i_{t+2}}(m),\cdots,\mathfrak{s}^\sigma_{i_{t+L-1}}(m)\}$.

An ($k+m$)-order partition $\mathcal{C}^\sigma(k+m)$ is generated by the $k$-pre-iteration of the boundary curves composing the $m$-order partition $\mathcal{C}^\sigma(m)$. The pre-iteration is given by the evolution operator $\mathcal{U}^{-k}$. This order-$(k+ m)$ partition is formed by the cells $\mathfrak{c}_i^{\sigma}(k+m)$.
Notice that a cell $\mathfrak{c}_i^{XY}(k+m)\in\; \mathcal{C}^{XY}(m+k)$, with $\mathcal{C}^{XY}(m+k)\equiv\; \mathcal{U}^{-k}(\mathcal{C}^{XY}(m))$ in an order-($k+ m$) partition represents points that follow a particular length-$L=k+m$ symbolic itinerary (or length-$L$ trajectory) in the order-$1$ partition. The probability measure of a length-$L$ itinerary $\mu(\{ \mathfrak{c}^\sigma_{i_{t-L}}(1),\mathfrak{c}^\sigma_{i_{t-L+1}}(1),\cdots,\mathfrak{c}^\sigma_{i_{t-1}}(1)\})$ is assumed to be equal to the probability measure of points in a cell of an order-$L$ partition and given by $\mu(\{\mathfrak{c}_i^{\sigma}(L)\})$, with $x_{t-L} \in \mathfrak{c}_i^\sigma(L)$. Many length-$L$ trajectories can follow the same itinerary. The probability $P$ calculated over the symbolic sequence of a length-$L$ itinerary along the order-$m$ partition is represented by $P(\{ \mathfrak{s}^\sigma_{i_{t-L}}(1),\mathfrak{s}^\sigma_{i_{t-L+1}}(1),\cdots,\mathfrak{s}^\sigma_{i_{t-1}}(1)\})$. If the partition is generating, then this probability is also equal to the probability of  points to belong to a cell $\mathfrak{s}_i(L)$ of an order $L$ partition and given by $P(\mathfrak{s}_i(L))$, with $i$ such that $\mathcal{T}(X_{-L})=\mathfrak{s}_i(L)$, where $\mathfrak{s}_i(L)$  is a length-$L$ symbolic sequence that gives the name of a cell in an order-$L$ partition. 

Thus, there are two ways of calculating probabilities. One based on the probability of the trajectory itineraries, which produce the probability measures $\mu$. The other based on the symbolic itinerary, which produces the probability measures $P$. There is however a fundamental difference between both probabilities. Whereas $\mu$ is calculated over a higher $(m+k)$-order partition with non-overlapping well defined cells, and therefore, it requires the use of a generating partition, $P$ refers to the probability of a symbolic sequence defined by the marginal lower $m$-order original partition. It therefore does not require that the higher-order partition is generating. In our practical numerical calculations, we adopt 
the probabilities $P$ to calculate our informational quantities. 

We assume in the following that the initial partition is order-1 ($m=1$), therefore, there is only one straight line in $\mathcal{L}_X(1)$ and one straight line in $\mathcal{L}_Y(1)$. Each of these symbolic itineraries along the order-1 original partition can be encoded by the symbolic name of a cell in an order-$L$ partition. 
An event in $\mathcal{C}^{XY}(1)$  is defined by trajectory points falling in $\mathfrak{c}_i^{XY}(1)\in\mathcal{C}^{XY}(1)$.

\subsection{A generating partition}

An order-$m+k$ partition $\mathcal{C}^\sigma(m+k)$  is generated from an order-m $\mathcal{C}^\sigma(m)$ by
\begin{equation}
\mathcal{C}^{\sigma}(m+k)=\mathcal{U}^{-k}(C^{\sigma}(m)).
\label{generating}
\end{equation}

The partition $\mathcal{C}^{\sigma}(m)$ is a "generating" partition if the all the cells in  $\mathcal{C}^{\sigma}(m+k)$ are non-overlapping and the union of the partition boundary curves in an order-$(m+1)$ partition restores the boundary curves of an order-$(m)$ partition.   
The higher order partitions shown in this work satisfy Eq. (\ref{generating}), however, higher-order cells do overlap. This overlapping is however minor to the orders treated here, although will affect the topological properties of the probabilistic space, this is not significant to all the results presented in this work, in particular to the nominal value of the CaMI quantity. This naturally will prevent the observed partitions to be generating. Consequently, this would have an impact if one would want to calculate CaMI for longer time-delays, in terms of the order-1 partition.  Not the case for the present work.

{{In practice, when dealing with experimental (with noise) or simulated time-series, as we will proceed in this work, the boundaries and the cells of higher order-$L$ partitions are determined from symbolic sequences of length $L$ created from an order-1 partition. Initial conditions generating a symbolic sequences of length $L$  in the $X_{-L}$  coordinate created from an order-1 partition (a binary partition) will belong to an order-$L$ marginal partition along this coordinate. Marginal and joint probabilities are also estimated from the probabilities of appearances of symbolic sequences by Eqs. (\ref{eq:ms3}) and (\ref{eq:ms4}), respectively. More details on how this is done including the composed subspace $Y_{-L}Y_{L}$ where joint events in the variable $Y$ are calculated and where marginal partitions will have an order of $2L$ can be seen in Sec. \ref{sec:topology_causal}. See also Eq. (\ref{eq:cami1}) and following explanations to understand how to calculate CaMI from the higher-order partitions, from symbolic sequences.}} 

\subsection{An example}

As an example of how trajectory points visit the partition with different orders and how this trajectory is symbolic encoded and probabilities are calculated, we consider a dynamic process along a 1D binary partition. Assume $x_t\in\mathcal{C}^X(1)$. Then, $\Phi^X_{-L}(t)=\{x_{t-L},x_{t-L+1},\cdots,x_{t-1}\}=X_{-L}(t)$ and $\Phi^X_{L}(t)=\{x_{t},x_{t+1},\cdots,x_{t+L-1}\}=X_{L}(t)$, and $S_{-L}^X(t,1)=\{\mathfrak{s}_{i_{t-L}}^X(1),$ $\mathfrak{s}_{i_{t-L+1}}^X(1),$ $\cdots,\mathfrak{s}_{i_{t-1}}^X(1)\}$ and 
$S_{L}^X(t,1)=\{\mathfrak{s}_{i_{t}}^X(1),$ $\mathfrak{s}_{i_{t+1}}^X(1),\cdots,\mathfrak{s}_{i_{t+L-1}}^X(1)\}$. Observing a trajectory composed by $\{X_{-L}(t),X_{L}(t)\}$ considering the partition $\mathcal{C}^X(1)$ allow us to conclude that the  trajectory of the system has visited a sequence of cells described by a sequence of symbols $\{S_{-L}^{XY}(t,1),S_{L}^{XY}(t,1)\}$.

Assuming $L=2$, it exists an order-2 partition generated by $\mathcal{U}^{-1}(\mathcal{C}^X(1))$ whose cells represent intervals where points within generate $S_{-2}^X(t,1)$ and $S_{2}^X(t,1)$.
Moreover, a cell in an order-2 partition is encoded by a symbol that represents the whole symbolic sequence of length-2 trajectories along the partition $\mathcal{C}^X(1)$, and therefore there exist $i$ such that $\mathfrak{s}_{i}^X(2)=S_{-2}(t,1)^X$ and there exist $j$ such that $\mathfrak{s}_{j}^X(2)=S_{2}(t,1)^X$. Consequently, $P(\mathfrak{s}_{i}^X(2))=P(S_{-2}(t,1)^X)$ and $P(\mathfrak{s}_{j}^X(2))=P(S_{2}(t,1)^X)$. 

In Fig. \ref{fig:diagram2}, we can see the relationship between a trajectory of length-4 in an order-1 and an order-2 partitions. This figure also illustrates how an order-2 partition is generated from an order-1 partition.

\begin{figure}[htb]
	\hspace{3pc}
	\includegraphics[scale=0.3]{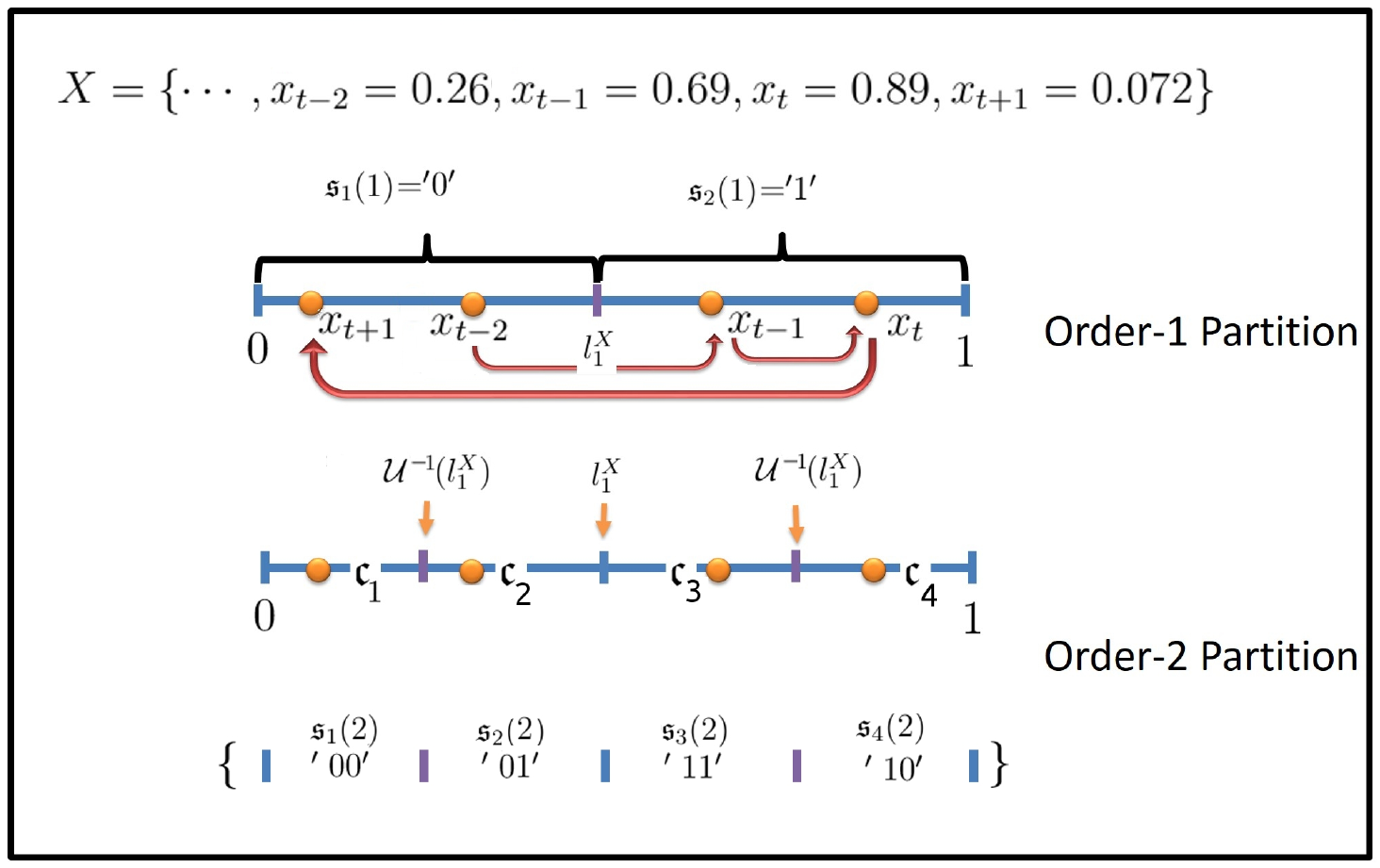}
	\caption{In this diagram a length-4 trajectory is plotted in the state space and two partitions are shown, one of order-1 defined by the line $l_1^X$ (and the borders of the state space) and another of order-2 defined by the union of $l_1^X$ with $\mathcal{U}^{-1}(l_1^X)$. The trajectory along the order-1 partition generates the symbolic itinerary $\{S_{-L}^X(1),S_{L}^X(1)\}=\{\mathfrak{s}_{t-2}='0',\mathfrak{s}_{t-1}='1',\mathfrak{s}_{t}='1',\mathfrak{s}_{t+1}='0'\}$ and along the order-2 partition generates the symbolic itinerary $\{S_{-L}^X(t,2),S_{L}^X(t,2)\}=\{\mathfrak{s}_{t-2}(2)='01',\mathfrak{s}_{t-1}(2)='11',\mathfrak{s}_{t}(2)='10',\mathfrak{s}_{t+1}(2)='00'\}$. Notice, however, that the first symbol in $\mathfrak{s}_{t}(2)='10'$ represents the present location in an order-1 partition and the second symbol represents the location of the first iteration. Therefore, the second symbol in $\mathfrak{s}_{t}(2)='10'$, represents the same first symbol in $\mathfrak{s}_{t+1}(2)='00'$}
	\label{fig:diagram2}
\end{figure}

\subsection{Probabilistic spaces and informational quantities} 

We now define some notations for the probabilities and informational quantities to simplify the exposition of our next derivations. 

The notations $P(X_L)$ or $P(Y_L)$ represent $P(\mathfrak{c}_i^{X_L}(L))$ or $P(\mathfrak{c}_i^{Y_L}(L)).$ So, $P(X_L)$ represents the probabilities of finding points in the cells $\mathfrak{c}_i^{X_L}(L)$ or similarly $P(X_L)=P(\mathfrak{s}_i^{X_L}(L))$. Therefore, the Shannon's entropy of length-$L$ symbolic sequences, represented by $H(X_L)$ or $H(X_{-L})$, is calculated by 

\begin{equation}\label{eq:ms1}
H(X_L)=-\sum_i P(\mathfrak{c}_i^{X_L}(L))\log(P(\mathfrak{c}_i^{X_L}(L))),
\end{equation}

So, if the "generating" property of the partition holds, entropies of length-$L$ trajectories along an order-1 partition space can be  calculated by the measure of the cells encoding symbolic sequences of length 1 appearing in the higher order-$L$ partitions. This approach is specially oriented for analytical derivations based on the study of networks of coupled dynamical systems. Otherwise, when dealing with time-series coming from experiments or numerical simulations, as it is the case of the present work, we calculate entropies considering the probabilities of length-$L$ symbolic sequences observed along the order-1 partition. So,
\begin{equation}\label{eq:ms3}
H(X_L)=-\sum_iP(\mathfrak{s}_i^{X_L}(L))\log(P(\mathfrak{s}_i^{X_L}(L))),
\end{equation}
where $\mathfrak{s}_i^{X_L}(L)$ represents a length-$L$ symbolic sequence. 

\begin{figure}[!ht]
\centering
\includegraphics[scale=0.25]{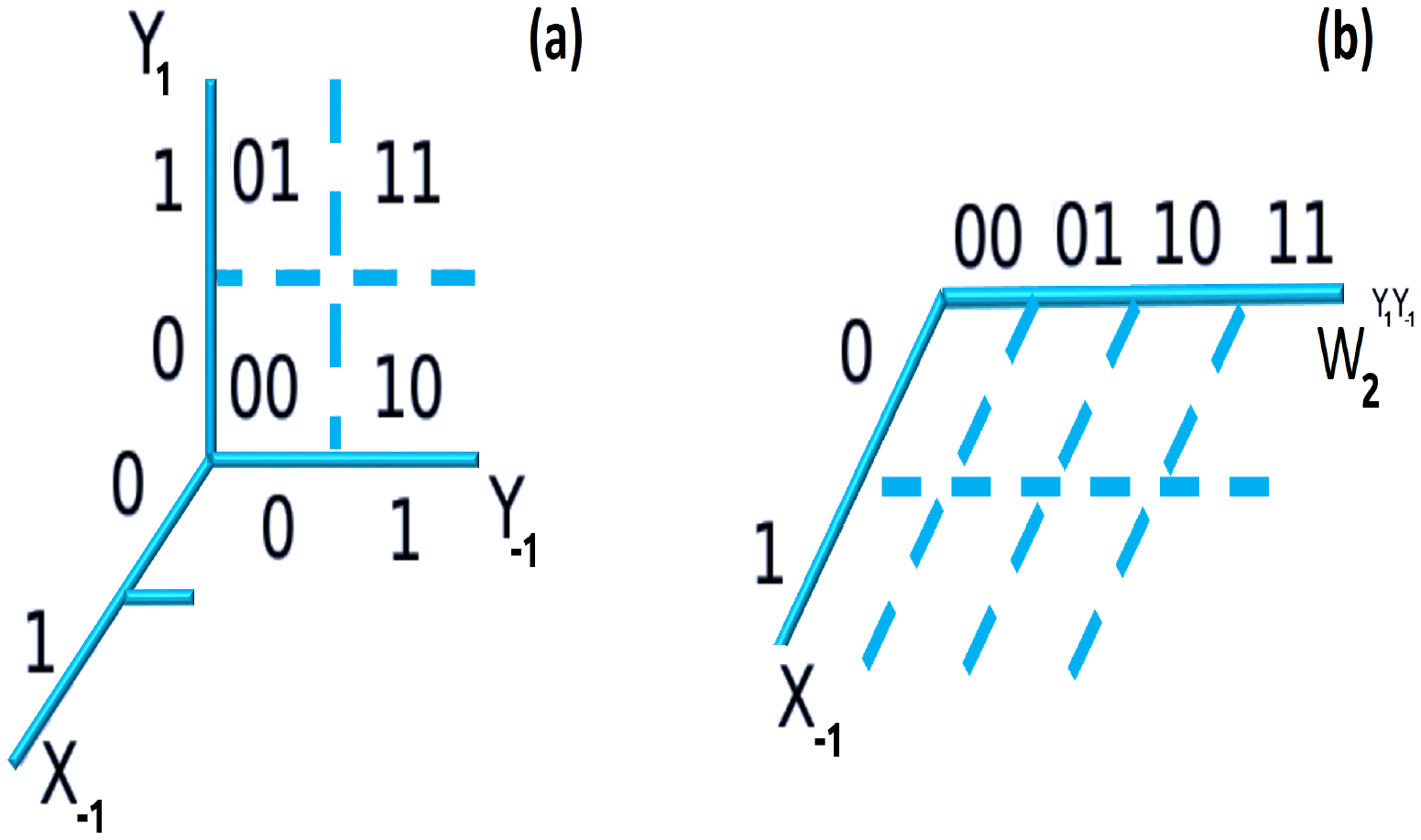}
\hspace{2pc}
\caption{(a) Visualization of the space $\Omega^{X_{-L}Y_{-L}Y_{L}}$ formed by the coordinates $(X_{-1}, Y_{-1},Y_1)$ and an order-1 partition in each subspace formed by pair of coordinates. (b) Visualization of the reduced-dimensional space with coordinates $(X_{-1},W_{2}^{Y_{-1}Y_1})$ and its asymmetric partition. Along $X_{-1}$ an order-1 partition is considered. Along $W_{2}^{Y_{-1}Y_1}$ an order-2 partition is considered.}
\label{fig:tri_dynamical}
\end{figure}

Let us now define a $2$-dimensional state space $\Omega^{X_{-L}Y_{-L}}\equiv [X_{-L}\times Y_{-L}]$ and with a trajectory $\Phi^{X_{-L}Y_{-L}}$ on it. 

The joint entropy of a composed space $H(X_L,Y_L)$ is then defined by 

\begin{equation}\label{eq:ms2}
H(X_L,Y_L)=-\sum P(\mathfrak{c}_i^{X_{L}Y_L}(L))\log(P(\mathfrak{c}_i^{X_{L}Y_{L}}(L))).
\end{equation}

Assuming that the generating property of the partitions do not hold, then the quantity in Eq. (\ref{eq:ms2}) is estimated by   
\begin{equation}\label{eq:ms4}
H(X_L,Y_L)=-\sum_iP(\mathfrak{s}_i^{X_L Y_L}(L))\log(P(\mathfrak{s}_i^{X_LY_L}(L))).
\end{equation}

By $P(X_L,Y_L)$, we represent the probability of a joint event calculated by $P(\mathfrak{c}_i^{X_L(L) Y_L(L)})$, and by $P(X_L | Y_L)$, we represent a conditional probability event representing the probability density 
of a point falling in the row $\mathfrak{c}_i^{Y_L(L)}$ and then being iterated to the column $\mathfrak{c}_i^{X_L(L)}$.

Extended $3$-dimensional spaces can be constructed by the composition of $\Omega^{X_{-L}Y_{-L}}$ with the 1-dimensional space representing the present of variable $Y$ observed in an order-$L$ partition, or a space constructed from the time-forward trajectory that visits an itinerary of length-$L$ in an order-1 partition. We represent this space by ($\Omega^{Y_L}$) or $X$ ($\Omega^{X_L}$). Notice that a point belonging to a partition in $\mathfrak{c}_i^{Y_L}(L)$ will produce an length-$L$ itinerary along the order-1 partition.   

We are interested in the spaces $\Omega^{X_{-L}Y_{-L}Y_{L}}$ or $\Omega^{Y_{-L}X_{-L}X_{L}}$ composed by the variables $\{X_{-L}(t),Y_{-L}(t),Y_{L}(t)\}$ or $\{Y_{-L}(t),X_{-L}(t),X_{L}(t)\}.$ It will be of further interest the spaces $\Omega^{Y_{-L}Y_{L}}$ and $\Omega^{X_{-L}X_{L}}$


Figure \ref{fig:tri_dynamical}(a) shows the space $\Omega^{X_{-L}Y_{-L}Y_{L}}$ formed by the time-delay coordinates $X_{-L}$, $Y_{-L}$ and time-forward coordinate $Y_{L}$, with $L=1$, and an order-1 partition in all subspaces defining our probabilistic space. The order-1 partition for the 2D space formed by ($Y_{-1}\times Y_1$) shows the symbolic names of columns, rows and the composed cells. 

Notice that the 2D space $\{Y_{-1},Y_1\}$ with an order-1 partition, where probabilities are calculated (Fig. \ref{fig:tri_dynamical}(a)), can be reduced to a 1D space $W_2^{Y_1Y_{-1}}$ with an order-2 partition (Fig. \ref{fig:tri_dynamical}(b)). A partition cell in the space $W_2^{Y_1Y_{-1}}$ represents point that are in $\mathfrak{c}_i^{Y_{-1}}(l)$ and move to $\mathfrak{c}_j(1)\in Y_{1}$, and therefore produce probabilities of the joint events $P(Y_{-1},Y_1)=P(Y_{-1})P(Y_{1}|Y_{-1})$. In a general situation, for an arbitrary $L$, probabilities in the space $\{X_{-L},Y_{-L}\}$ and $\{Y_{-L},Y_{L}\}$ could be calculated over an order-$L$ partition on each subspace. The reduced probabilistic space would be composed by a coordinate $X_{-L}$ where probabilities are calculated over an order-$L$ partition and the coordinate $W^{Y_{-L}Y_L}_{2L}$ where probabilities would be calculated over an order-$2L$ partition. A cell in $W^{Y_{-L}Y_L}_{2L}$ would represent joint events $P(Y_{-L},Y_L)=P(Y_{-L})P(Y_{L}|Y_{-L})$.

Notice that one can consider subspaces $X_{-L}$ and $Y_{-L}$ with an order-$L$ partition each, and the subspace $Y_J$ with an order-$J$ partition, with $J\neq L$, composing the space $\Omega^{X_{-L}Y_{-L}Y_J}$. Then, the reduced space $W$ would have a probabilistic space formed by a partition of order $(J+L)$. 

\section{The topology of causality}\label{sec:topology_causal}
\subsection{Generating higher-order partitions}

We consider two non-coupled Logistic maps ($\alpha=0$ in Eq. (\ref{coupled_maps_network})), represented by $X$ and $Y$, to illustrate how we construct our partitions. Setting $l_1^X=0.5$, if $x_i\leq l_1^X$ then $\mathcal{T}(x_i)=0$ (and $\mathfrak{s}_i^X(1)=``0"$), and if $x_i > l_1^X$ then $\mathcal{T}(x_i)=1$ (and $\mathfrak{s}_i^X(1)=``1"$). Applying these rules for a trajectory of this uncoupled system, we generate Fig. \ref{fig:pre_column}(a). In Fig. \ref{fig:pre_column}(b), we show in green two columns of the order-2 partition obtained by $\mathcal{U}^{-1}(l_1^X)$. Setting $l_1^Y=0.5$, therefore $\mathcal{L}^{XY}=\{l_1^X,l_1^Y\},$ we generate Fig. \ref{fig:pre_column}(c).  The same coloured regions in this figure represent cells in an order-3 partition created by $\mathcal{U}^{-2}(\mathcal{C}^{XY}(1))$.  The order-3 partition has columns and rows enclosed by straight lines $\mathcal{L}_X(3)=\mathcal{U}^{-2}(l_1^X(1))$ and $\mathcal{L}_Y(3)=\mathcal{U}^{-2}(l_1^Y(1))$ respectively, forming the partition $\mathcal{C}^{XY}(3)=\mathcal{U}^{-2}(C^{XY}(1))$.  In this case, each column and row have boundaries that describe a generating partition of one Logistic map.

\begin{figure}[!ht]
\centering
\includegraphics[scale=0.35]{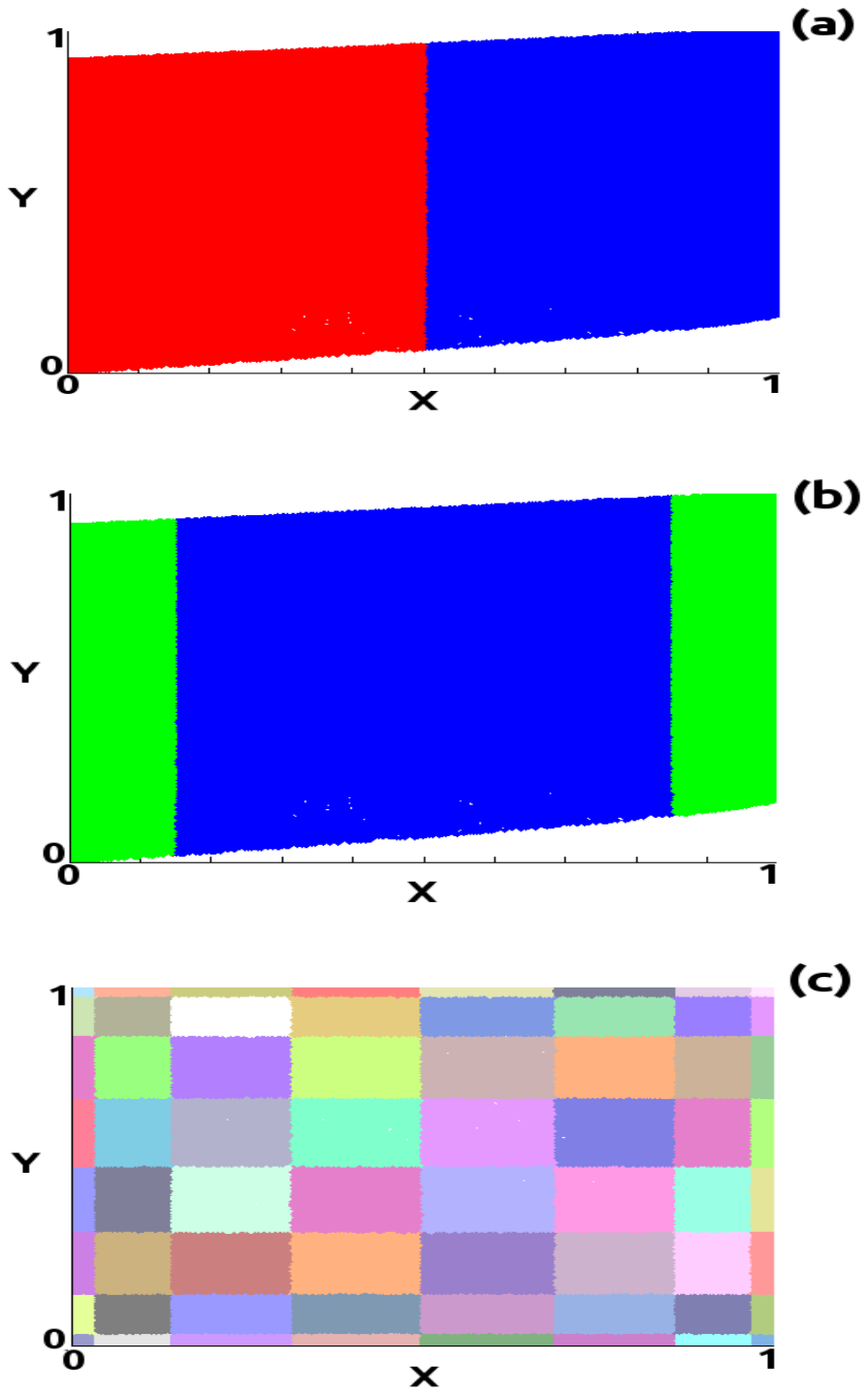}
\caption{Panel (a) shows two columns with name $\mathfrak{s}_1(1)=``0"$ in red and $\mathfrak{s}_2(1)=``1"$ in blue, for the order-1 partition in $X$. Panel (b) shows in green two columns from an order-2 partition in $X$ with names $\mathfrak{s}_1(2)=``00"$ and $\mathfrak{s}_4(2)=``10"$. Panel (c) shows, by the regions of the same color, cells $\mathfrak{c}_i^{XY}(3)$ of an order-3 partition.}
\label{fig:pre_column}
\end{figure}
In practice, we do not make pre-iterations of the partition lines to determine the higher-order partitions. Once we choose $\mathcal{L}_X$ and $\mathcal{L}_Y$, the rows, columns, and cells of higher-order partitions are visualised by the colours of points that encode a particular symbolic sequence, using the following algorithm. Given a trajectory, we construct the length-$L$ segments of it 
$\Phi^{XY}_{L}(n)=\{(x^1_{n},x^2_{n}), \ldots, (x^1_{n+L},x^2_{n_L})\}$, and whose symbolic sequence is represented by $S_{L}^{XY}(n,1)=\{\mathfrak{s}_{n}^X(1),\mathfrak{s}_{n+1}^X(1), \ldots, \mathfrak{s}_{n+L-1}^X(1) \bullet \mathfrak{s}_{n}^Y(1),\mathfrak{s}_{n+1}^Y(1), \ldots, \mathfrak{s}_{n+L-1}^Y(1)\}$. This symbolic sequence  is then encoded into an integer number that is used in the palete of colours to set the colour of the point $(x^1_{n},x^2_{n})$ that will produce the length-$L$ symbolic sequence $S_{L}^{XY}(n,1)$. Points will belong to the same column (row) if their symbolic sequence  $S_{L}^{X}(n,1)$ ($S_{L}^{Y}(n,1)$) is the same, and will belong to the same cell if their 
symbolic sequence   $S_{L}^{XY}(n,1)$ is the same. To set the palete of colours, we produce an integer for the colour of the point $\mathbf{\Phi}_L(n) = (x^1_n,x^2_n)$ of an order-$k$ partition generated from an order-1 partition, 
using the following encoding rule 

\begin{equation}
colour(\mathbf{\Phi}_L(n)) = \theta_x(n)*2^k+\theta_y(n), 
\label{encoding1}
\end{equation} where $\theta_x(n)=\sum_{i=1}^L \mathfrak{s}_{n+1-i}^X(1)2^{L-i}$ and $\theta_y(n)=\sum_{i=1}^{2L} \mathfrak{s}_{n+1-i}^Y(1)2^{2L-i}$.  Then, colours are randomly assigned to each of these integer numbers.
 
\subsection{Understanding the arrow of influence}

We now show how the topological properties of higher-order partitions change according to the coupling strength $\alpha$ between 2 or more coupled systems as in Eq. \eqref{coupled_maps_network}, and how these topological asymmetries can be used to determine the arrow of influence in these systems. 

The  symmetry in the structure of the partition in Fig. \ref{fig:pre_column}(c) reflects the fact that the two systems are not coupled. 
Imagine that an observer measure an event in the variable $Y$ at a time $n$: $Y(n)=x_n^2=0.5$. Applying $\mathcal{U}^{-k}(l_1^Y)$, for any $k$, will create always a vertical line stretching from  0 to 1, meaning that an observation in $Y$ at time $n$ cannot be used to localize the state of the variable $X$ at time $n-k$. The consequence, as we will show next is that there is no flow of information from $X$ to $Y$. The contrary is also true, i.e., one can also use similar arguments to conclude that there is no flow of information from $Y$ to $X$. 


For a coupled system in a master (node $X$) and a slave (node $Y$) configuration, assuming a coupling strength of $\alpha=0.09$, we have created partitions of different orders (from order-1 to order-5) and shown in Fig. \ref{fig:order_partitions}. One can see how the increase of the order increases the topological complexity of the partitions, for example going from Fig. \ref{fig:order_partitions}(a) to Fig. \ref{fig:order_partitions}(d). 
Paying attention to the higher order rows, defined by the enclosure of $\mathcal{U}^{-k}(l_1^Y)$ along the $Y$ variable, in Fig. \ref{fig:order_partitions}(e-h), and the higher-order columns, defined by enclosure of $\mathcal{U}^{-k}(l_1^X),$  we can observe how they are not enclosed any longer by straight lines. This asymmetry is the consequence of $X$ driving $Y$.

\begin{figure}[ht]
\centering
\includegraphics[scale=0.4]{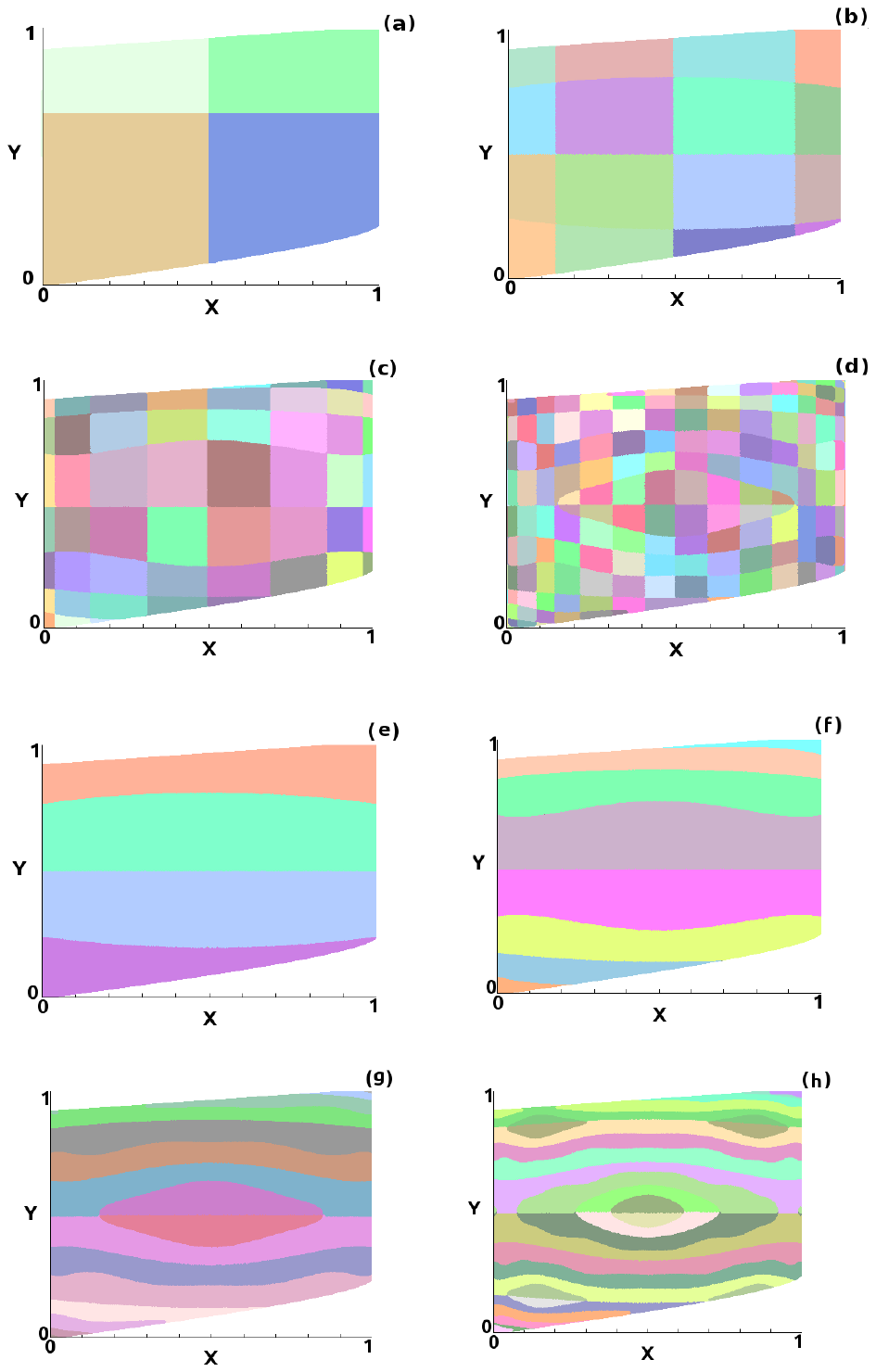}
\caption{Panel (a) shows an order-1 partition $\mathcal{C}^{XY}_{1},$ each coloured region represents a cell. Panel (b) shows the order-2 partition given by $\mathcal{C}^{XY}_1=\mathcal{U}^{-1}(\mathcal{C}^{XY}(1))$. Panels (c-d) show the corresponding order-3 and order-4 partitions, generated by $\mathcal{U}^{-2}(\mathcal{C}^{XY}(1))$ and $\mathcal{U}^{-3}(\mathcal{C}^{XY}(1))$ respectively, coloured regions represent the cells in the partition. Panels (e-g) show $\mathcal{U}^{-1}(l^{Y}(1)),\mathcal{U}^{-2}(l^{Y}(1))$ and $\mathcal{U}^{-3}(l^{Y}(1))$ respectively, and each coloured "horizontal" stripe represents a higher-order row. Finally panel (h) shows only the higher-order rows of an order-5 partition, enclosed by $\mathcal{U}^{-4}(l^{Y}(1))$.}
\label{fig:order_partitions}
\end{figure}

We now want to analyse the different topological features of the higher-order partitions when we consider different orders in $X$ and $Y$. For that we produce Fig. \ref{fig:CaMI_logistics} obtained from two Logistic maps coupled in a master ($X$) and slave ($Y$) configuration with a coupling strength $\alpha=0.09$. We have selected two orders for our partitions in $X$ and $Y$: 2 and 5. 
Figure \ref{fig:CaMI_logistics} (a) shows the different cells (same colour region) of a partition created by the intersection of an order 2 partition in $X$ and order 5 partition in $Y$. Figure \ref{fig:CaMI_logistics}(b) shows the different cells of a partition of an order 2 in $Y$ and an order 5 in $X$. The asymmetry in Fig. \ref{fig:CaMI_logistics}(a) indicates that the system has an arrow of influence $X \to Y.$

\begin{figure}[ht]
\centering
\includegraphics[scale=0.35]{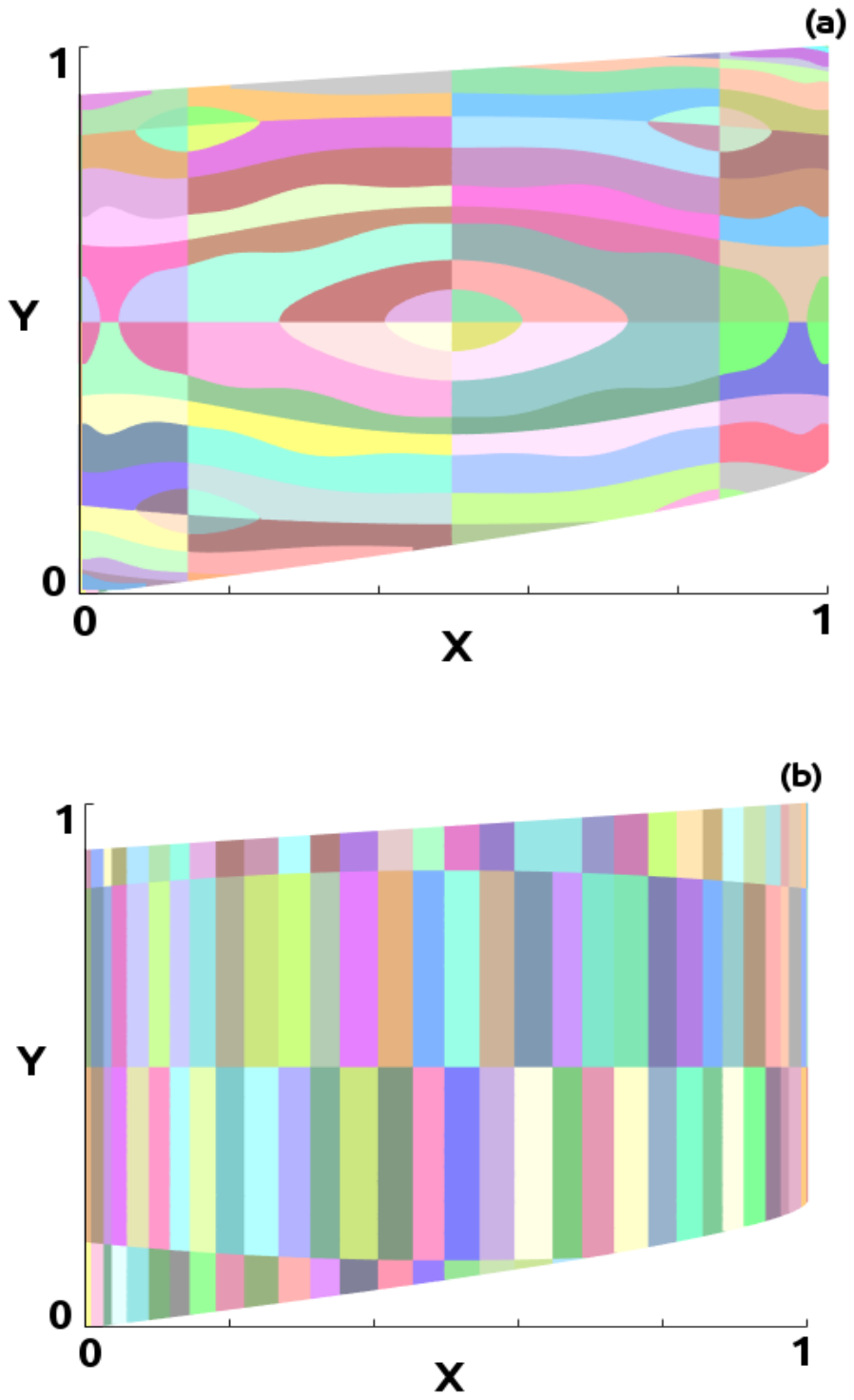}
\caption{Panel (a) shows the partition of order 5 in $Y$ and order 2 in $X$. Panel (b) shows the partition of order 2 in $Y$ and order 5 in $X$. }
\label{fig:CaMI_logistics}
\end{figure}
\subsection{Local mutual information}

Pointwise mutual information (PMI) is a probabilistic measure of the amount of information that two different random variables posses locally between them. Given a particular partition, the PMI only takes into consideration the information computed over a single cell and not over the entire set of cells as the MI. PMI is defined as 

\begin{equation}\label{eq:pmi}
PMI(i,j)=H^{i}_{X}+H^{j}_{Y}-H^{i,j}_{XY},
\end{equation}
where: $H^{i}_{X}=-P(\mathfrak{s}_i^X)\log(P(\mathfrak{s}_i^X)),$ $H^{j}_{Y}=-P(\mathfrak{s}_j^Y)\log(P(\mathfrak{s}_j^Y)),$ and $H^{i,j}_{XY}=-P(\mathfrak{s}_{k}^{XY})$ $\log(P(\mathfrak{s}_{k}^{XY})),$ with $k$ representing the cell formed by the overlapping of the higher-order row $i$ with the higher-order column $j$. Therefore, MI is just the average of PMI over cells of the partition. In the following, we consider a normalized variant of PMI, named normalized Pointwise Mutual Information (nPMI), introduced in Ref.\cite{pmi}, and defined as:

\begin{equation}
nPMI=-\frac{PMI(i,j)}{\log \left(P(\mathfrak{s}_{k}^{XY})\right)}.
\end{equation}

The advantage of the nPMI over PMI is the reduction of the sensitivity of the measure to short time-series.

\begin{figure}[htb]
\centering
\includegraphics[scale=0.25]{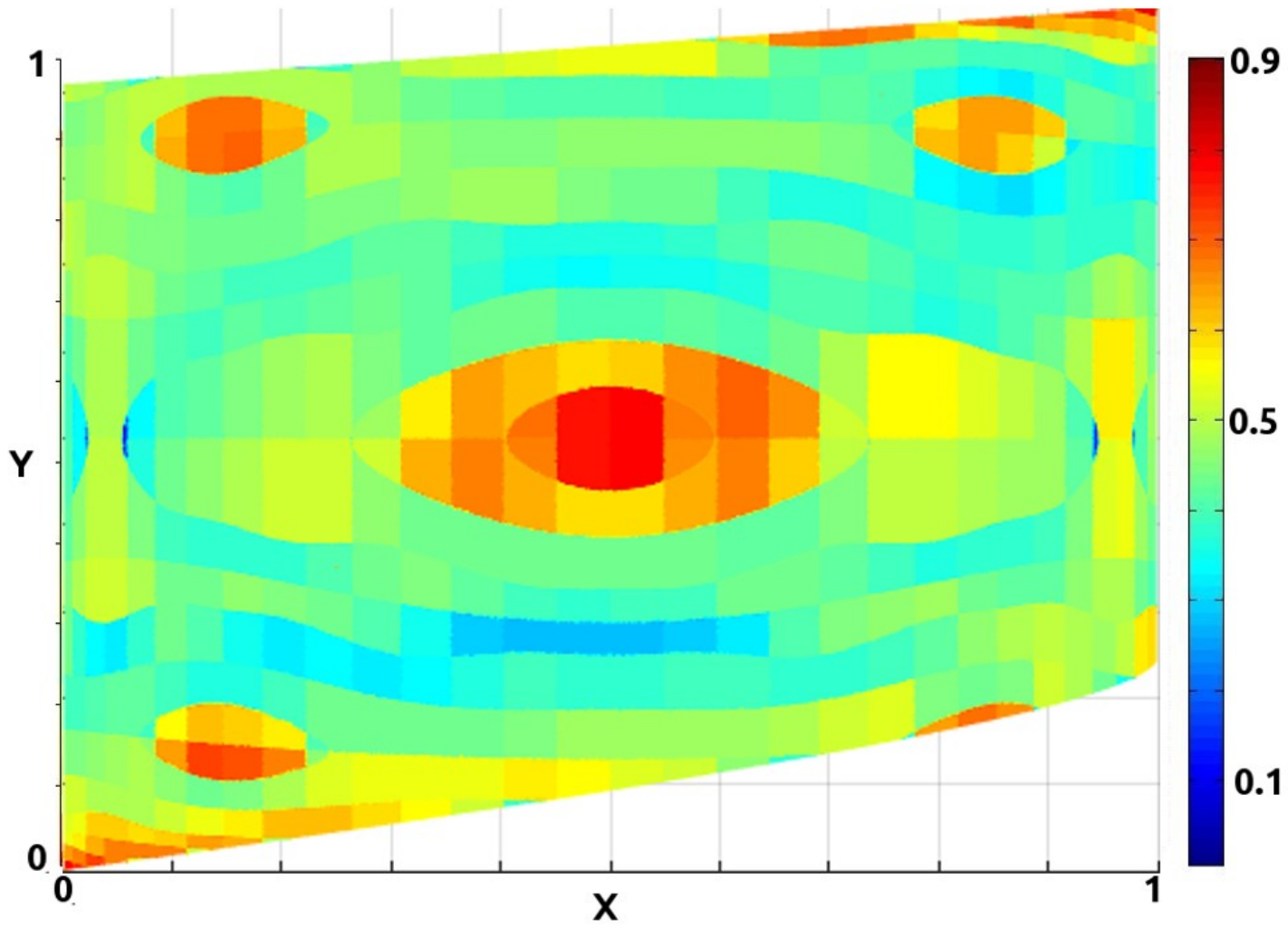}
\caption{nPMI for a directed coupled Logistic map with a coupling strength of 0.09. It can be observed how the nPMI is higher inside the bubbles. Minimum and maximal values of nPMI are -5.12 and 4.12 respectively. These values were normalized to fit in the interval $[a,b]$, with $a=0$ and $b=1$. The equation for the normalization is 
$y=(b-a)*(\mbox{(nPMI)}-\min{(nMPI)})/(\max{(nMPI)}-\min{(nMPI)})+a$.}
\label{fig:npmi}
\end{figure}

Using nPMI in the partitions considered in Fig. \ref{fig:CaMI_logistics}, we can calculate the amount of information exchanged between variables $X$ and $Y$. The nMPI values for each cell represent the contribution of specific symbolic sequences (time nature of causality), or of particular variables domains (spatial nature of causality) contributing to the information transferred from $X$ to $Y$. 


Figure \ref{fig:npmi} shows the nPMI for two directed coupled Logistic maps for an order-5 partition $\mathcal{C}^{XY}$ as the order of the partition along the variable $Y$ in Fig. \ref{fig:CaMI_logistics}(a). The surprising fact is that nPMI is larger for a special union of cells enclosed by one of the solutions of $\mathcal{U}^{-3}(l_1^Y)$ that forms closed curves. We call these special union of cells, causal bubbles, from its closed graphical representation. We can see that the bubbles are areas containing trajectory points responsible for a large amount of information exchanged between $X$ and $Y,$ and as explained in the following, a consequence of the fact that $X\to Y.$

\subsection{Causal bubbles}\label{sec:bubbles}

Figure \ref{fig:buble_lines} shows an illustration of how these bubbles are created. The partition line $l_1^Y$ represented by the black dashed line in panel (a) is iterated once producing the red curves such as the ones showed in panel (b), and eventually after $2L-1$ backwards iterations these curves form a closed contour as is shown in panel (c) by the closed red curves.
\begin{figure}[htb]
\centering
\includegraphics[scale=0.35]{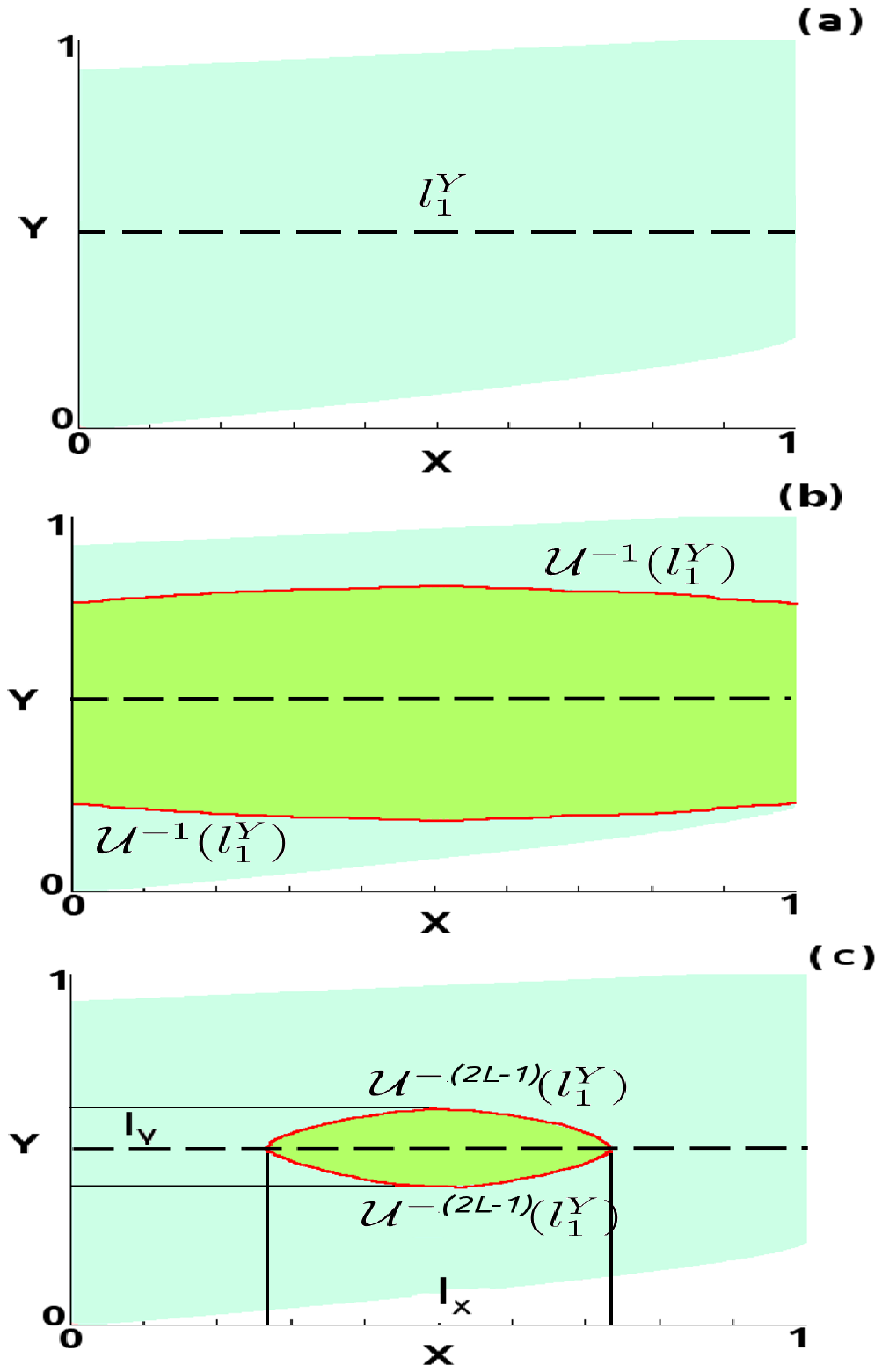}
\caption{(a) Dashed black line represents $l_1^Y$. (b) The red curves are obtained by one backward iteration of $l_1^Y.$ (c) The closed red curve is obtained by $2L-1$ backward iterations of $l_1^Y$.}
\label{fig:buble_lines}
\end{figure}

 Assume that the future of the observed variable $y_{t+L-1}$, $L$ iterations forward in $y$, has a value that lays exactly at the partition line, i.e. $y_{t+L-1}\in l_1^Y$. The variable to be predicted has an arbitrary value at $t+L-1$, so $x_{t+L-1}$ can assume any value in the state space. Then, after one backward iteration, $y_{t+L-2}\in \mathcal{U}^{-1}(l_1^Y)$ is located on the red lines in panel (b). After $2L-1$ backward iterations $y_{t-L}$ has a position along the red closed curve in panel (c), enclosing a bubble area.

Therefore, the first observation in the variable $y$ at time $t+L-1$ cannot tell anything about the position of the variable $x_{t+L-1}$. Assuming the observer has full knowledge of the dynamical equations, it makes one observation at $y$ obtaining the value $y_{t+L-1}$, which in this imaginary example lays at $l_1^Y$. The smart observer uses the knowledge of the dynamics and makes $2L-1$ backward iteration of $l_1^Y.$ The observer will conclude that, if $y_{t+L-1}\in l_1^Y$, then $y_{t-L}\in I_Y$ and $x_{t-L}\in I_X,$ where $I_X$ is the $x$-interval enclosed by the red curve, panel(c). So, by making one observation of the future of $y$ and using the knowledge of the dynamics ({{which can be obtained by inspection of the time-series or from modelling the system}}), the observer can better localize the state of the past variable $x$,  if $X\to Y.$ Moreover, if more observations are done (from the future to the past) in $y,$ more likely the observer is to improve its knowledge about the location of $x_{t-L}$ by doing similar analysis. 

If there is no physical connection between system $X$ and system $Y,$ and no flow of information from $X$ to $Y,$ the bubbles are not formed, and therefore, one cannot localize the position of the variable $X$ by observing $Y.$

{{It is worth mentioning that the studied system is non-invertible, and therefore, if future of $Y$ depends on the past $X$ (as one would naturally conclude if doing the Granger analysis by constructing a model from the data), it is not necessarily obvious that the past of $X$ can be predicted by the future of $Y$. The causal bubble however demonstrates that this is indeed true. }}

\section{Causal mutual information (CaMI)}

We are now ready to define a new informational quantity named \textit{Causal Mutual Information} from $X$ to $Y$ (CaMI$_{X\to Y}$) as the mutual information between joint events in $X_{-L}$ and the set composed by the joint events of $Y_{-L}$ and $Y_{L}$ as

\begin{equation}\label{eq:cami1}
\mbox{CaMI}_{X \to Y}=MI(X_{-L};Y_{-L},Y_{L})=MI(X_{-L};W^{Y_{L}Y_{-L}}_{2L}).
\end{equation}
Analogously, CaMI$_{Y \to X}=MI(Y_{-L};W^{X_{L}X_{-L}}_{2L})$. {{So, in practice, CaMI$_{X \to Y}$ is calculated by computing the mutual information between symbolic sequences of length $L$ in the variable $X$ and symbolic sequences of length $2L$ in the variable $Y$.}}

Notice that CaMI is not permutable since, CaMI$_{X \to Y}\neq \mbox{CaMI}_{Y \to X}$. As we can see, CaMI$_{X\to Y}$ is the Mutual information between trajectory points in the subspace $X_L$ and the subspace $W^{Y_{L}Y_{-L}}_{2L}$ and that measures the amount of information between longer time-series of past, present and future of $Y$ and shorter time-series of the past of $X$. The  fundamental idea behind the reason for us to propose CaMI as a measure of causality is that if there is a flow of information from $X$ to $Y$, then longer observations in the $Y$ variable can be used to predict the past of states of $X$. CaMI is also a quantity that measures the total amount of information extracted from one variable by observing another variable, not only the shared amount (non causal, measured by the Mutual Information) but also the transmitted amount (causal, measured by the Transfer Entropy).  

Considering the coupled system and the same partition used to create Fig. \ref{fig:CaMI_logistics}(a), the magnitude of the computed CaMI$_{X \to Y}$ is 0.17. Considering the coupled system and the same partition used to create Fig. \ref{fig:CaMI_logistics}(b), the magnitude of the computed CaMI$_{Y \to X}$ is 0.04. The difference in the CaMI's magnitudes, meaning CaMI$_{X \to Y}$-CaMI$_{Y \to X}>0$, and the asymmetry in Fig. \ref{fig:CaMI_logistics}(a) with the existence of the causal bubbles indicate the presence of a system whose direction of causality is given by $X \to Y,$ and therefore, there is a flow of influence from $X$ to $Y$. As we shall see, for this situation, CaMI$_{Y \to X}$=$MI(X_{-L};Y_{-L})$, meaning that the variables share some common, non causal information.   
 
 
Notice that the PMI as defined in Eq. \eqref{eq:pmi} is just one of the terms considered in the calculation of CaMI. 
 

\subsection{Transfer Entropy and Causal Mutual Information}

Having two random processes $X$ and $Y$, the amount of information transferred from process $X$ to $Y$ ($X\rightarrow Y$) can be quantified by the Transfer Entropy \cite{schreiber}, defined as:
\begin{equation}\label{transfer_entropy}
T_{X\rightarrow Y} = H\left( Y_{L} \mid Y_{-L}\right) - H\left( Y_{L} \mid Y_{-L}, X_{-L}\right),
\end{equation}
where $H(X)$ is the Shannon's entropy of $X$, $H(X_{L} \mid Y_{-L})$ is the knowledge (reduce of entropy) of process $X$ from time $t$ to $t+L-1$ if the past of process $Y$ from $t-1$ to $t-L$ is known, and  $H\left( Y_{L} \mid Y_{-L}, X_{-L}\right)$ represents the knowledge of the process $Y$ from time $t$ to $t+L-1$, if the past of $X$ and $Y$ in the interval from $t-1$ to $t-L$ is known. 
Transfer entropy was shown to be related  \cite{newton2016transfer,liu2012relationship} to directed information, a sort of a cumulative version of transfer entropy. The later quantity is being currently considered as an appropriate measure to deal with channels of communication with fee-back, describing channels where the output is fed back to the input.

We can express Eq. \eqref{transfer_entropy} as a function of joint entropies and not conditional one using te chain rule for entropy and Bayes theorem:
\begin{equation}\label{transfer_joint}
\begin{split}
T_{X\rightarrow Y} = H(Y_{L},Y_{-L})-H(Y_{-L})-\\H(Y_{-L},Y_{L},X_{-L}) +
H(Y_{-L},X_{-L})
\end{split}
\end{equation}

But notice that CaMI can also be written as 
\begin{equation}
\label{cami-new}
\begin{split}
\mbox{CaMI}_{X\to Y}  = MI(X_{-L};Y_{L},Y_{-L})  = \\ H(X_{-L})+H(Y_L, Y_{-L})  - H(X_{-L},Y_{L},Y_{-L}).
\end{split}
\end{equation}

Finally, comparing  Eq. (\ref{cami-new}) with (\ref{transfer_joint}), we conclude that 
\begin{equation}\label{eq:TvsCami}
\mbox{CaMI}_{X\to Y} = T_{X\rightarrow Y}  + MI(X_{-L};Y_{-L}),
\end{equation}
where $MI(X_{-L};Y_{-L})=H(X_{-L})+ H(Y_{-L})-H(X_{-L},Y_{-L})$ is the mutual information of the system composed by $X$ and $Y$. Both quantities are shown over a Venn diagram in Fig. \ref{fig:CaMI_Venn} for their comparison. One can see that CaMI carries more information about the considered variables than transfer entropy. CaMI$_{X \to Y}$ represents the amount of information exchanged between $X$ and $Y$ (provided by the term $MI(X_{-L};Y_{-L})$) and the transfer entropy from $X$ to $Y$. Whereas  $MI(X_{-L};Y_{-L})$ measures how much the observation of a length-$L$ trajectory along the variable $X$ (or $Y$)  can be predicted by observations of a length-$L$ trajectory of the variable $Y$ (or $X$), the transfer entropy from 
$X$ to $Y$ measures how much one can predict from the past state of the $X$ by making observations of the past, present, and future states of the variable $Y$.   
\begin{figure}[htb]
\centering
\includegraphics[scale=0.3]{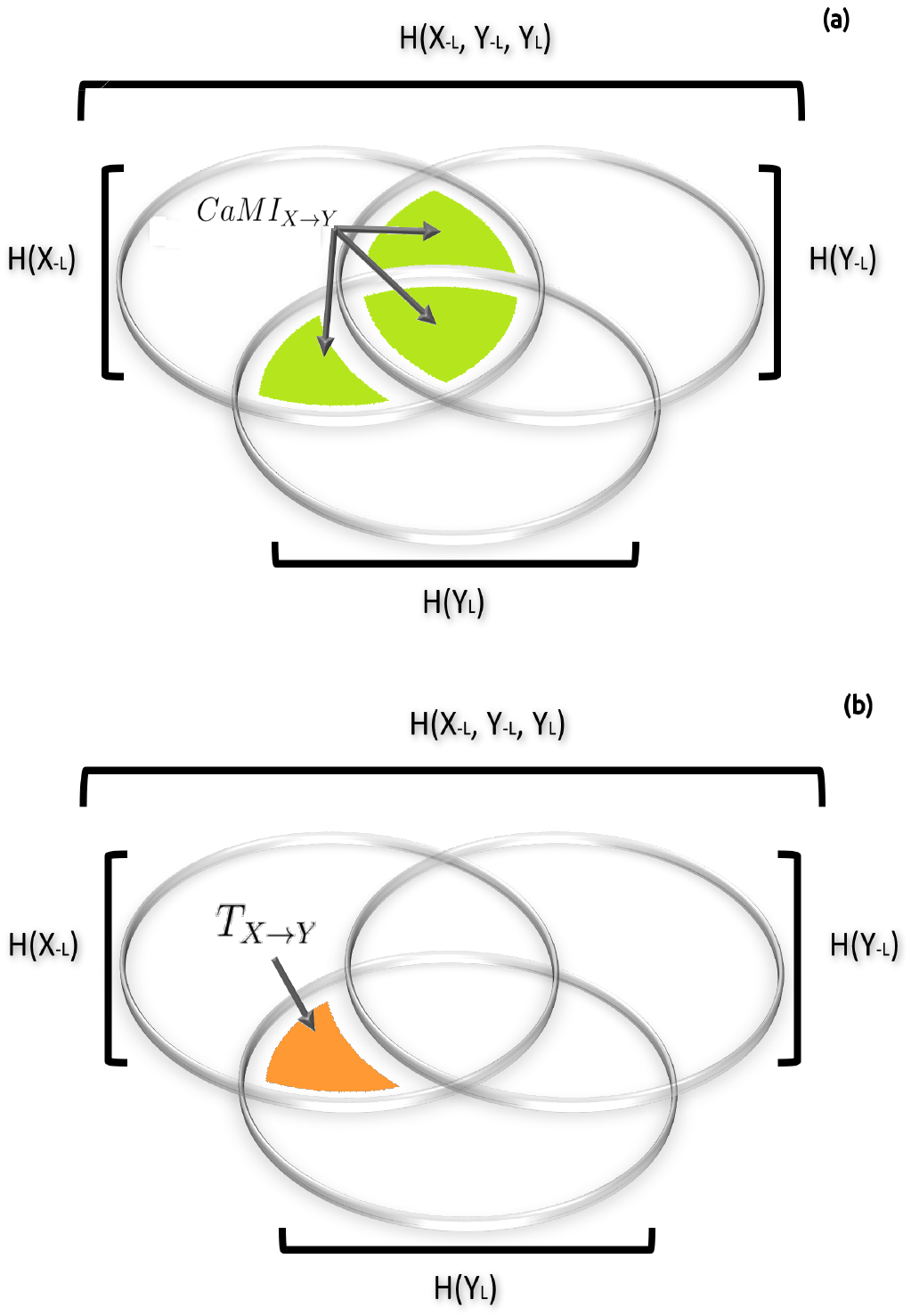}
\caption{Representation over a Venn diagram of the CaMI. Panel (a): Causal Mutual Information, CaMI$_{X\rightarrow Y}$. Panel (b): Transfer entropy, $T_{X\rightarrow Y}$.}
\label{fig:CaMI_Venn}
\end{figure}
One important fact to notice is that since $MI(X_{-L};Y_{-L}) = MI(Y_{-L};X_{-L})$, the directionality index defined by $T_{X\to Y}-T_{Y \to X}$, and therefore representing the net flow of information between both variables, can be calculated by
\begin{equation}
\begin{array}{l}
DI = T_{X\rightarrow Y}-T_{Y\rightarrow X}=\mbox{CaMI}_{X\rightarrow Y}-\mbox{CaMI}_{Y\rightarrow X}.
\end{array}
\end{equation}

As another remark, notice that transfer entropy is defined as the conditional mutual information and therefore, $T(X\to Y)=MI(X_{-L};Y_{L}|Y_{-L})$, see Ref. \cite{palus}. Recalling that CaMI$_{X \to  Y}=MI(X_{-L};(Y_{-L},Y_{L}))$, it is easy to see that to define CaMI we have replaced the conditional probabilities in the transfer entropy $T_{X \to Y}$ to joint probability ones in CaMI$_{X \to Y}$. 

\section{Higher-resolutions initial marginal partitions}

We have previously seem that the probabilities of length-$L$ symbolic sequences representing trajectory points following an itinerary along order-1 marginal partitions can be used to calculate CaMI, and therefore, be used to detect the causal direction of the flow of information. We have seen that  
the topology of a higher-order 2D asymmetric partition, where each coordinate has different orders, can be used to determine the arrow of influence between two variables. The higher-order partitions were generated out of order-1 marginal partitions. However, in certain practical situations, for example in stochastic or experimental systems, higher-order partitions generated out of lower-order partitions should be expected to produce no discernible topological set that could orient one to the correct direction of the flow of information. It is thus interesting to verify whether and for which boundary conditions a marginal partition with $N_X$ boundary lines along the variable  $X$ ($\mathcal{L}_X(m)$) and a marginal partition with $N_Y$ boundary lines along the variable $Y$ ($\mathcal{L}_Y(2m)$) could be used to estimate a physically consistent CaMI$_{X \to Y}$. Simply put, CaMI$_{X \to Y}$ calculated in this way would be obtained by measuring the mutual information between lower-resolution observations along the variable $X$ and higher-resolutions observations along the variable $Y$. 
  
We study the mutual information of a system (as the one shown in Fig. \ref{fig:2logistic_maps_bidirected}), with a coupling strength $\alpha=\{0,0.05,0.1\}$ and $\beta=0.1-\alpha$. In Fig. \ref{fig:MI_colormaps}, we show by colours the values of the MI calculated considering Eqs. \eqref{eq:ms1} and \eqref{eq:ms2} between variables $X$ and $Y$ for a partition $\mathcal{C}^X$ for $X$ and  $\mathcal{C}^Y$ for $Y$, with different number of columns and rows, respectively. The number of N$_X$ and N$_Y$ are shown in the axis of Fig. \ref{fig:MI_colormaps}. Recall that CaMI$_{X\to Y}$ is just the mutual information between variables $X$ and $Y$ where the partition of $X$ has order $L$ and the partition of $Y$ has order $2L$. In here we want to test the plausible idea that causality can also be detected when the variables are observed with different spatial resolutions.  We, therefore, want to test whether MI is capable of detecting the direction for the flow of information when a probabilistic space has an arbitrary number of equal rectangular areas.

In Figs \ref{fig:MI_colormaps}(a)-(c), if the partition in $X$ has the same number of cells than the partition in $Y$, MI grows with the growing of the number of cells. As expected, one can see that if the flow of information is from $X\to Y$ (as in panel (a)), then partitions with more cells in $Y$ than in $X$ produce larger MI (or CaMI) than partition with less cells in $Y$ than in $X$.

\begin{figure}[htb]
	\centering
	\includegraphics[height=1.5cm]{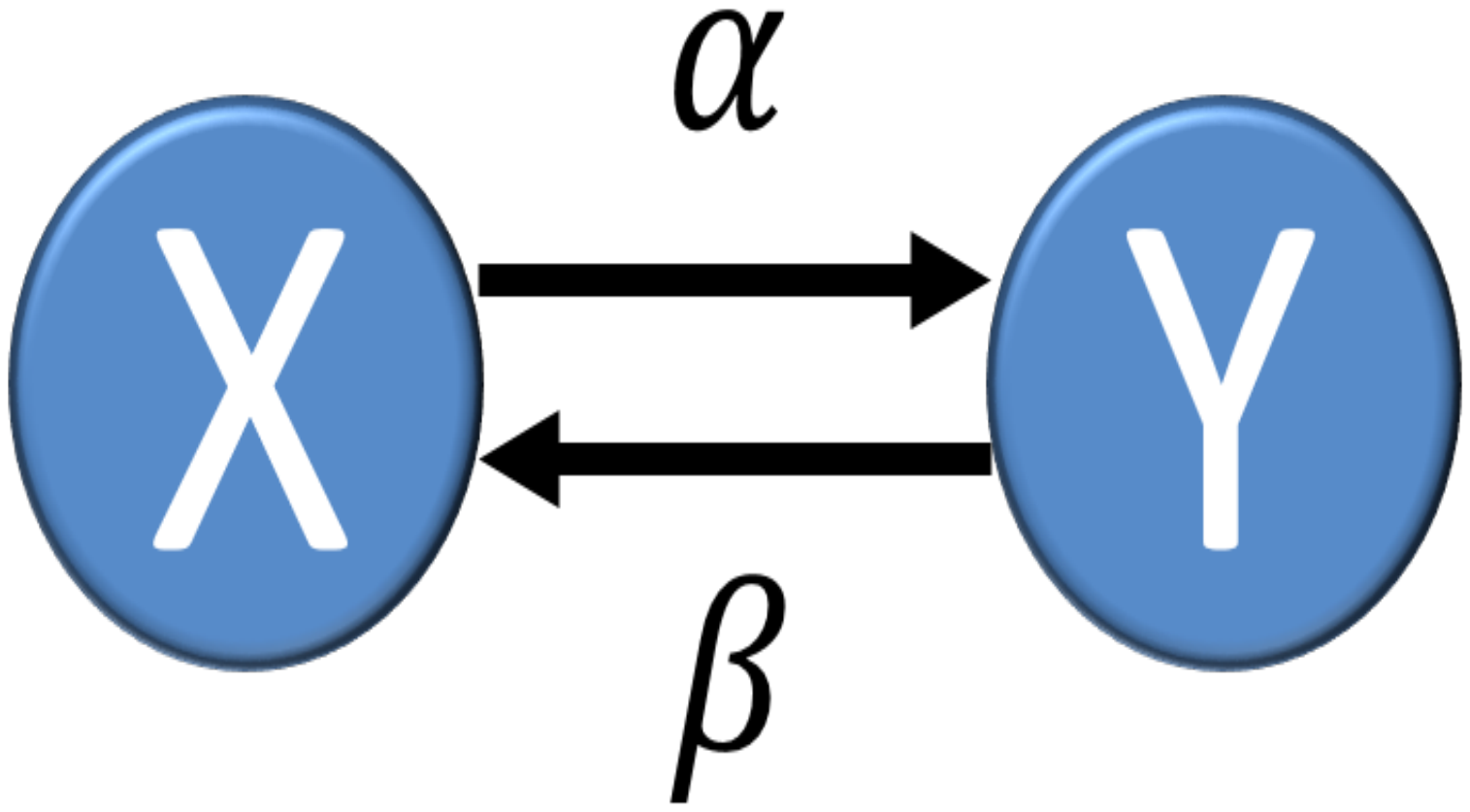}
	\caption{Two Logistic maps bi-directionally coupled . The coupling strengths $\alpha$ and $\beta$ are related by $\beta=0.1-\alpha$.}
	\label{fig:2logistic_maps_bidirected}
\end{figure}
Surprisingly, Fig. \ref{fig:MI_colormaps}(a) shows a novel feature for the MI of coupled systems. If the flow of information goes from $X \to Y,$ then given a partition in $Y$ with a particular number of cells (i.e., resolution), the value of MI obtained is roughly invariant for the chosen resolution in $X$. This implies that the amount of information one can realize from $X$ by making measurements in $Y$ is almost solely dependent on the resolution of the observation in $Y$, and there is a sufficient large amount of number of cells in $Y$. In Fig. \ref{fig:MI_colormaps}(b) the flow of information goes from $Y$ to $X$ and therefore, for a sufficiently large number of cells in $X$, the information that one can deduce from $Y$ by measuring $X$ almost solely depends on the resolution of $X$. Finally, in Fig. \ref{fig:MI_colormaps} (c) for the bidirectionally coupled system, with equal coupling strengths, the values for MI will depend on both resolution of variables $X$ and $Y$ in a complementary way, i.e. if the sum of the number of rows and columns is maintained, MI remains roughly invariant. In fact, the relationship is described by a diagonal hyperbola. {{This phenomenon can also be used to detect the directionality of the coupling.}} 

\begin{figure}[htb]
\centering
\includegraphics[scale=0.25]{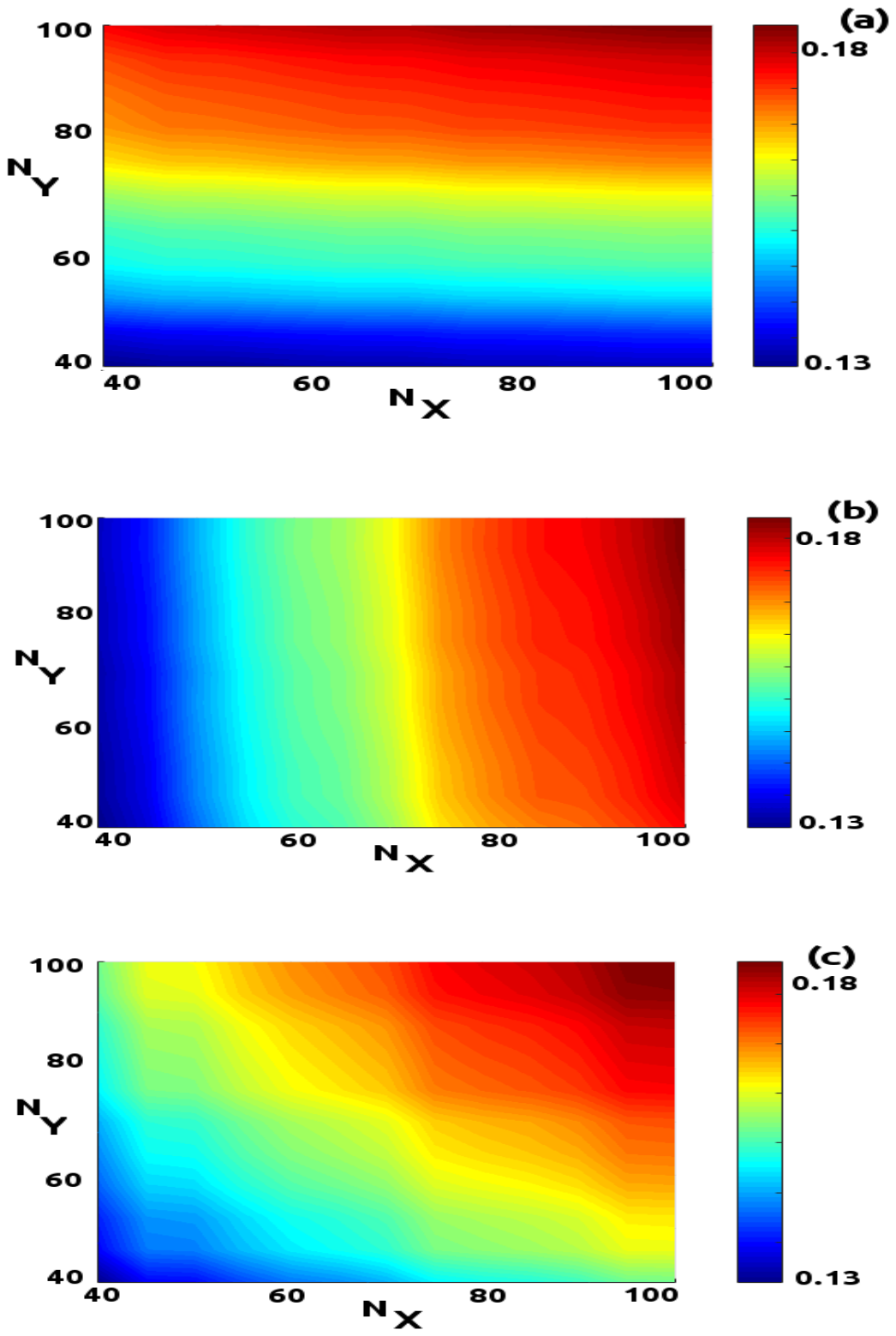}
\caption{Mutual information of coupled Logistic maps computed on an asymmetric partition with different number of rows and columns.
The horizontal and vertical axis show the different amount of rows and columns of the partition. Panel (a) shows the result for  $\alpha=0.1$ and $\beta=0$, so information flows from $X$ to $Y$. Panel (b) the results for $\alpha=0$ and $\beta=0.1$, so information flows from $X$ to $Y$. Panel(c) shows the result for $\alpha=0.05$ and $\beta=0.05$}
\label{fig:MI_colormaps}
\end{figure}

\vspace{-0.5cm}

\section{Conclusion}


In this work, we have investigated the spatio-temporal properties of causality, causality meaning the study of the arrow of influence between two systems.  The spatio-temporal nature of causality can be exploited to detect the arrow of influence from $X$ to $Y$, either by considering shorter time-series of $X$ and longer time-series of $Y$, an approach that explores the time nature of causality, or  lower precision measured time-series in $X$ and higher precision measured time-series in $Y$, an approach that explores the spatial nature of causality.  Thus, this work shows that causality can be detected not only by the analyses of the topological properties of higher-order partitions generated by lower-order marginal partitions (the "space" property of causality), but also  by considering the probabilities on these partitions, reflecting the density of trajectories of a given time-length 
(the "time" property of causality).  To apply this abstract notion of causality into a quantitative approach, we have introduced a new informational quantity, the causal mutual information,  CaMI$_{X \to Y}$ that measures the total amount of information being transmitted from $X$ to $Y$, the information shared between both variables and that can be used to predict the present state of $X$ by observations in $Y$ (i.e., the Mutual Information), and the causal information directed transmitted from $X$ to $Y$, which can be used to predict the past states of $X$ by observations of the past and future states of $Y$ (i.e., the Transfer Entropy). 

Since CaMI does not require the calculations of conditional probabilities, but rather only joint probabilities, the probabilistic spaces involved in its calculation can be lower-dimensional, {{enabling a quick estimation of TE. This property is well wished for causal analysis of large complex systems such as the brain, or for technological applications that employ TE, for example,}} the recently proposed brain-based cryptography (see Ref. \cite{szmoski}). Also, less data is required for the determination of causality, since the probability space can be constructed according to the available data.

Another important result of this work was to show that measuring a driven variable with finer resolution than that used to observe the driving variable allows us to obtain more information about the driving system, but not the other way around. Increasing the resolution of observation of the driving variable brings no additional information about the driven variable. This observation could also be exploited to detect directionality in networked systems.  

A potential advantage of our approach is that even thought the analysis of causality is bivariate,
employing two observables taken from two subsystems $X$ and $Y$ in a larger system, the topological properties of the constructed probabilistic space can potentially discern whether information is being sent physically from $X$ to $Y$, or whether it is being mediated by other subsystems and variables (in this case, there is no physical connection between $X$ and $Y$).  This special property of the probabilistic space shown to exist to fully deterministic systems to detect causality allows one to detect direct or mediate effects without the need to calculate multivariate conditional probabilities, from which one can detect direct or mediated influences in stochastic systems Ref. \cite{runge,bollt}, an approach suitable for both dynamical and stochastic systems, but that however requires the use of large dimensional probabilistic spaces. In the multivariate approach to detect causality, the conditional probabilities of multivariate variables need to take into consideration the influence of co-founders \cite{mangiarotti2015}, entities that mediates the transfer of information from $X$ and $Y$ systems.  In fact, our approach was recently tested to infer the topology and the synaptic nature (either chemical or electrical) of complex neural networks. It was shown \cite{Borges2017} that CaMI can be used to successfully infer the topology of the neural network with no mistakes, and also discern about the nature of the connections, even when the network is in the presence of both dynamical and observational additive Gaussian noise, and even when only observational time-series based on local averages are available.   

\section*{Acknowledgments}
Authors acknowledge EPSRC Ref: EP/I032606/1 grant.

\vspace{-0.5cm}

\section*{Bibliography}
%





\end{document}